\documentclass[structabstract]{aa}  
\usepackage{graphicx}
\usepackage{txfonts}
\usepackage{natbib}
\usepackage{rotating}

\newcommand{\ergs}{erg\ s$^{-1}$}

\newcommand{\xmm}{{\sc XMM}\emph{-Newton}}
\newcommand{\ros}{\emph{{\sc ROSAT}}}
\newcommand{\ch}{\emph{{\sc Chandra}}}

\newcommand{\loglxlb}{$\log[L_{\rm X}/L_{\rm BOL}]$}

\begin{document}
   \title{Hot stars observed by \xmm\ \\I. The catalog and the properties of OB stars\thanks{Based on observations collected with XMM-Newton, an ESA Science Mission with instruments and contributions directly funded by ESA Member States and the USA (NASA), and accessed via the 2XMMi and XMM slew survey catalogs.}}
\authorrunning{Ya\"el Naz\'e}
\titlerunning{Hot stars observed by \xmm\ I.}
   \author{Ya\"el Naz\'e
          \thanks{Postdoctoral Researcher FNRS}\fnmsep\thanks{Visiting astronomer, UNAM-Morelos (Mexico)}
          }

   \institute{GAPHE, D\'epartement AGO, Universit\'e de Li\`ege, All\'ee du 6 Ao\^ut 17, Bat. B5C, B4000-Li\`ege, Belgium\\
              \email{naze@astro.ulg.ac.be}
             }


 
  \abstract
   {} 
   {Following the advent of increasingly sensitive X-ray observatories, deep observations of early-type stars became possible. However, the results for only a few objects or clusters have until now been reported and there has been no large survey comparable to that based upon the ROSAT All-Sky Survey (RASS). }
   {A limited survey of X-ray sources, consisting of all public \xmm\ observations (2XMMi) and slew survey data (XMMSL1), is now available. The X-ray counterparts to hot, massive stars have been searched for in these catalogs. }
   {About 300 OB stars were detected with \xmm. Half of them were bright enough for a spectral analysis to be possible, and we make available the detailed spectral properties that were derived. The X-ray spectra of O stars are represented well by low ($<$1keV) temperature components and seem to indicate that an absorption column is present in addition to the interstellar contribution. The X-ray fluxes are well correlated with the bolometric fluxes, with a scatter comparable to that of the RASS studies and thus larger than found previously with \xmm\ for some individual clusters. These results contrast with those of B stars that exhibit a large scatter in the $L_{\rm X}-L_{\rm BOL}$ relation, no additional absorption being found, and the fits indicate a plasma at higher temperatures. Variability (either within one exposure or between multiple exposures) was also investigated whenever possible: short-term variations are far more rare than long-term ones (the former affects a few percent of the sample, while the latter concerns between one third and two thirds of the sources). }
   {This paper presents the results of the first high-sensitivity investigation of the overall high-energy properties of a sizable sample of hot stars.}


   \maketitle
%

\section{Introduction}

Soon after the discovery of X-ray emission from hot stars 30 years ago, it was proposed that a correlation exists between the X-ray and bolometric luminosities \citep[e.g.][]{har79,pal81}, of the form $L_{\rm X}^{\rm unabs}\sim 10^{-7} \times L_{\rm bol}$. This relation probably reflects the dependence of the X-ray emission on the properties of the stellar winds, which are in turn strongly linked to the total luminosity of the star, since these winds are radiation-driven. Knowing and understanding this $L_{\rm X}-L_{\rm BOL}$ relation is thus of prime importance.

\citet{ber97} investigated this scaling law for a large sample of objects (237 OB stars). Their study relied on data from the \ros\ All-Sky Survey (RASS), which provided a rather homogeneous dataset, and represented the first large-scale survey of massive stars in the X-ray range. The derived $L_{\rm X}-L_{\rm BOL}$ relation exhibits a large scatter ($\sigma$ of 0.4 on a logarithmic scale, or a factor 2.5), and breaks down below $L_{\rm bol}\sim10^{38}$ \ergs, i.e., for mid and late B stars. 

In recent years, new studies investigating this so-called `canonical' relation have been performed, using detailed observations of rich open clusters and associations, notably NGC 6231 (\xmm, \citealt{san06}), Carina OB1 (\xmm, \citealt{ant08}), Westerlund 2 (\ch, \citealt{naz08a}), and Cyg OB2 (\ch, \citealt{alb07}). For the clusters with the most well constrained stellar content (NGC 6231 and Carina OB1), the derived $L_{\rm X}-L_{\rm bol}$ relations are far tighter (dispersions of only 40\%) than in the RASS analysis. 

The difference between the RASS and \xmm\ results may be caused by two factors. First, the datasets and data handlings are not directly comparable: on the one hand, the RASS study is based on a large sample of stars with approximately known properties (count rates, spectral type); on the other hand, the recent studies have considered a small sample of well-known stars with precisely known properties (derived from X-ray spectra and optical monitoring). Second,  the nature of the samples also differ: mixed field and cluster stars on the one hand, and a homogeneous stellar population from a single cluster on the other. Metallicity and age could indeed affect the star's X-ray properties. 

To determine which factor is the most important, I decided to investigate the high-energy properties of a large sample of hot stars, in a similar way as \citet{ber97} but by employing the higher sensitivity and resolution of \xmm. At the present time, two \xmm\ surveys are available: the 2XMMi catalog and the XMM slew survey (XMMSL1) catalog.

In Sect. 2, I describe the main properties of the 2XMMi catalog, in Sect. 3 the sample of hot stars detected in the 2XMMi catalog, and in Sect. 4 the sample of hot stars detected in the XMMSL1 catalog. Sections 5 and 6 present the discussion and conclusion, respectively.

\section{The 2XMMi catalog}

The \xmm\ observatory was launched in December 1999 \citep{jan01}. Its main instruments are three non-dispersive cameras, EPIC MOS1, MOS2, and pn \citep{str01,tur01}, and two grating spectrographs, RGS1 and 2 \citep{den01}. The EPIC cameras are sensitive to the 0.15--12.\,keV band, whereas the RGS studies instead the 0.35--2.5\,keV band. While the RGS provides the highest spectral resolution ($R$=300 at 1\,keV), medium-resolution spectroscopy ($R$=10 at 1\,keV) can also be obtained using the intrinsic energy resolution of the CCDs that make up the EPIC cameras. In addition, time series can be derived from EPIC data, enabling variability studies of the detected sources.

Since pointed observations often detect many serendipitous sources in addition to the main target, the XMM Survey Science Centre (SSC) consortium compiled a catalog of serendipitous X-ray sources using 4117 \xmm\ archival datasets. This was developed to help fully exploit the capabilities of \xmm, by means of a dedicated, homogeneous processing of all available data. The second main release of this catalog, called the 2XMM catalog, was published on-line in 2007 \citep{wat09} and an incremental version (2XMMi), more complete, was made available in the following year. A slim version of the 2XMMi catalog, containing the most important information, can be queried using Vizier \footnote{http://vizier.u-strasbg.fr/viz-bin/VizieR?-source=IX/40} while the full version can be downloaded from the Vizier ftp site \footnote{ftp://cdsarc.u-strasbg.fr/pub/cats/IX/40}. Detailed information (images, spectra, time series) as well as a user guide are available at the SSC website\footnote{http://amwdb.u-strasbg.fr/2xmmi/catexpert and http://xmmssc-www.star.le.ac.uk/Catalogue/2XMMi/UserGuide\_xmmcat.html respectively}.  

In this catalog, each source has a unique identifier based upon the IAU naming convention: it starts by 2XMM or 2XMMi, followed by the truncated position of the source at equinox 2000.00 (JHHMMSS.S$\pm$DDMMSS). If the source was detected in several exposures, additional entries are available, each being identified by a unique identifier (DETID) preceding the IAU name. More than 300 columns follow the name. They list the details of the observations (observation identifier, filters used, exposure times,...), as well as the basic source properties (count rates, hardness ratios) in each instrument and for several energy bands. Fluxes are also available but are not considered here because the count-rate-to-flux conversion used in the catalog considers a power-law spectrum typical of background AGNs, which indeed does not apply to hot stars. In this paper, the count rate in the 0.5-4.5\,keV energy band (aka ``band 9") is preferentially considered since this rather broad band was used for a separate run of the SAS analysis tasks, which ensured a better handling of the error values. Fortunately, it is also where hot stars emit most of their X-ray flux and the signal-to-noise ratio reaches its maximum. The four hardness ratios used in the 2XMMi catalog are defined as $HR=\frac{H-S}{H+S}$, where S=0.2--0.5\,keV and H=0.5--1.0\,keV for $HR_1$, S=0.5--1.0\,keV and H=1.0--2.0\,keV for $HR_2$, S=1.0--2.0\,keV and H=2.0--4.5\,keV for $HR_3$, and S=2.0--4.5\,keV and H=4.5--12.0\,keV for $HR_4$. Undetermined values are quoted as ``NULL'', which is the case for non-detections in one instrument (either because it was not turned on or because the source fell on a CCD gap). Time series and spectra are extracted for each detection of a given source if it presents at least 500 EPIC counts.

It must be underlined that this catalog should not be considered as a complete, homogeneous all-sky survey. First, the entire set of the regions observed by \xmm\ cover only 420 square degrees, or 1\% of the entire sky, and this coverage is very patchy. Second, these regions were not randomly chosen, but requested by different PIs for their specific needs, which means that the catalog is certainly biased: for example, about 65\% of the observed regions have large Galactic latitudes $|b|>20^{\circ}$ because of the strong interest in extragalactic fields. Finally, the exposure times range from shorter than 1000s to 130\,000s, and therefore the detection limit varies greatly from one area to another. However, as imperfect as it might be, it nonetheless constitutes the most sensitive X-ray catalog available at the present time in the observed regions, surpassing the RASS in every way. With $\sim$220\,000 unique detections, the 2XMMi catalog is the largest X-ray catalog ever produced: it contains twice as many discrete sources than previous surveys, notably the RASS.

\section{Finding hot stars in the 2XMMi catalog}

To find the hot stars detected by \xmm, it is necessary to correlate the 2XMMi catalog with a catalog of hot stars. To this aim, the Reed catalog of hot stars \citep{ree03} was chosen because it is one of the most complete for this type of objects. The catalog version of January 2009 was kindly provided by its author. It contains $>$19\,000 objects, mostly O and B stars, plus a few emission-line stars (WRs, T Tauri, HAe/Be,...). 

The correlation between these two catalogs was measured twice, once using the preferred ``source name'' in the Reed catalog and a second time using the coordinates provided by Reed (whenever Simbad was unable to resolve the provided name). The search radius was fixed to 5". This value represents a compromise. On the one hand, the \xmm\ PSF has a FWHM of $\sim$5" on-axis but this greatly increases towards the edges of the field-of-view. The position uncertainty is expected to be a fraction of the PSF, and 96\% of the detections indeed have a positional error of $<$5". On the other hand, residual shifts can exist between X-ray and optical observations and the optical coordinates are certainly not 100\% accurate.

If two (or more) X-ray sources were detected within the chosen correlation radius, the closest one was kept only if the second closest source was more distant by at least 1" (to avoid spurious multiple detections). In addition, extended sources were discarded, since these objects can either be true diffuse X-ray sources, sub-areas of extended emission, or spurious, combined detections of several stars in a dense cluster. High-mass X-ray binaries recorded in Simbad were also discarded, since this paper considers only the intrinsic emission of hot stars, and not their interaction with compact objects. Finally, a few detected sources were found to be listed twice in the Reed catalog: in these cases, only the smallest number used as an identifier in the Reed tables was listed in the correlation catalog\footnote{This is the case of ALS9458=ALS4879, 9465=4887, 9468=4889, 9477=4894, 9500=4912, 9506=4914, 9511=4918, 9526=4923, 9536=4929, 9537=4930, 9604=5025, 9609=5039, 9621=5046, 15858=1855, 21108=11417.}.

The 2XMMi catalog provides a ``quality flag" $S$, which is set to be zero if the X-ray source has no problem and to 1--4 if it might be spurious. Experience showed that hot stars in clusters are generally detected with a non-zero quality flag, because of the presence of numerous neighbours. To be as complete as possible, no selection was made based on the quality flag and all detections were kept. However, the higher quality objects are separated from the others in the discussion and tables provided below. To check whether spurious associations were a problem, the approach of \citet{ber96} was used. The number of background objects is expected to scale with the number of detected hot stars, the area of detection (5" radius), and the X-ray source density (about 500 src per square degree, compared to the 1.5 src per square degree of the RASS): for our data, only one spurious source is expected. Inspecting all bright sources (i.e., those with available light curves and spectra) identified only one really spurious detection, associated with ALS 4440 (=HD316464), which is part of the PSF wing of the bright X-ray binary XTE J1751--305. 

The result of the correlation between the Reed and 2XMMi catalogs is presented in Table 1 (available electronically at the CDS). The first column provides the data used to query the catalog, i.e., either the prefered name in the Reed catalog or the Reed coordinates. It is preceded by a letter in case two stars correspond to the same X-ray source. The second column lists the Reed number of the star (ALS \# in Simbad), while the third one provides the usual name of the source (HD, BD, CPD,... identifiers). The fourth column indicates the spectral type of the source, which was chosen to be the most recent from the Reed catalog, except for the Wolf-Rayet stars (WRs) where the classification of the VIIth WR catalog \citep{van01} was used. If no type is available in the Reed catalog, the spectral type from Simbad is reproduced here, preceded by a $\sim$ to easily distinguish these sources. The fifth column corresponds to a flag set to be Y if the source was detected as a binary. A star was classified as binary if either the spectral type available in the Reed catalog indicates the presence of a companion, or if the star is a known binary in the 9th Binary catalog \citep{pou04} and/or in \citet{gie03}. The next few columns indicate the V magnitude, B$-$V colour, interstellar absorption column, and bolometric flux. The photometric data correspond to the most recent values in the Reed catalog or, if unavailable, to the Simbad values (again flagged with a $\sim$). For the WRs, the interstellar absorption column and the bolometric fluxes were taken from \citet{osk05} for WN stars. For O and B stars, these characteristics were calculated whenever a precise spectral type (i.e., not only `O star' or `B star') and magnitude values were known. The interstellar column was derived from Bohlin's formula ($N_H=5.8\times 10^{21}\times E(B-V)$\,cm$^{-2}$, \citealt{boh78}), where the colour excesses were calculated from the difference between the observed colours and the instrinsic ones. Because of the scarcity of accurate distances, bolometric fluxes were preferred to bolometric luminosities. They were calculated using the usual formulae, yielding $\log(f_{\rm BOL})=-4.61-\frac{V-3.1\times E(B-V)+BC}{2.5}$, where $f_{\rm BOL}$ is in erg\,cm$^{-2}$\,s$^{-1}$. Intrinsic colours and bolometric corrections (BC) were taken, for the considered spectral types, from \citet{mar06} for O stars and \citet{sch82} for B stars. In the case of binaries, the intrinsic properties (colour, BC) of the primary star were used. The subsequent columns correspond to the X-ray source properties, i.e., the distance between the X-ray source and its optical counterpart, the official IAU name of that counterpart, list of the filters used for each EPIC instrument (pn, MOS1, and MOS2 in this order), the total count rates and their errors in the 0.5--4.5\,keV band for each instrument (with pattern 0--4 for pn and 0--12 for MOS), the four hardness ratios for each instrument, the total detection likelihood, the overall quality flag (see previous paragraph), two variability flags (see Sect. 3.1.2), the number of individual detections, the unabsorbed X-ray flux in the 0.5--10.\,keV band (see Sect. 3.2), and finally the $\log L_{\rm X}/L_{\rm BOL}$ value.

If several observations were available for a given source, a mean count rate was calculated for each instrument+filter configuration. Error bars were compared for different filter choices of a given instrument; the instrumental configuration corresponding to the smallest error bar was considered as the most suitable one and reproduced in Table 1.  



Note for individual objects: ALS 4440 is spurious (see above); the counterpart of ALS 4592 (=CPD\,$-$24$^{\circ}$6140) is real but affected by straylight, so that its properties might not be totally uncontaminated; ALS 20163 (=West 1 30) shares its X-ray counterpart with a sgBe star.

\subsection{Basic properties of the detected objects}

From the correlation performed above, 310 stars (with a known O, B, or WR spectral type) were found to have an \xmm\ counterpart within 5''. Of these, 133 have a zero quality flag and correspond to 132 unique 2XMMi sources (i.e., two sources are found within the PSF of one X-ray counterpart). The remaining 177 objects are considered as potentially spurious within the 2XMMi catalog; they correspond to 173 individual 2XMMi sources, i.e., there are four pairs of stars sharing the same X-ray counterpart. 

The biases inherent to the generation of this 2XMMi catalog have already been pointed out in Sect. 2 and the conclusions found from the performed correlation should indeed be interpreted with caution. Nevertheless, I examined the basic properties of the detected objects. While they may not be totally representative of the entire population of hot stars, these detections certainly represent the largest sample of such stars studied by modern X-ray observatories. Note that the remainder of the paper focuses mainly on the detailed properties of OB stars, the spectral analysis of WR stars being deferred to a future paper.

   \begin{figure*}
   \centering
   \includegraphics[width=9cm]{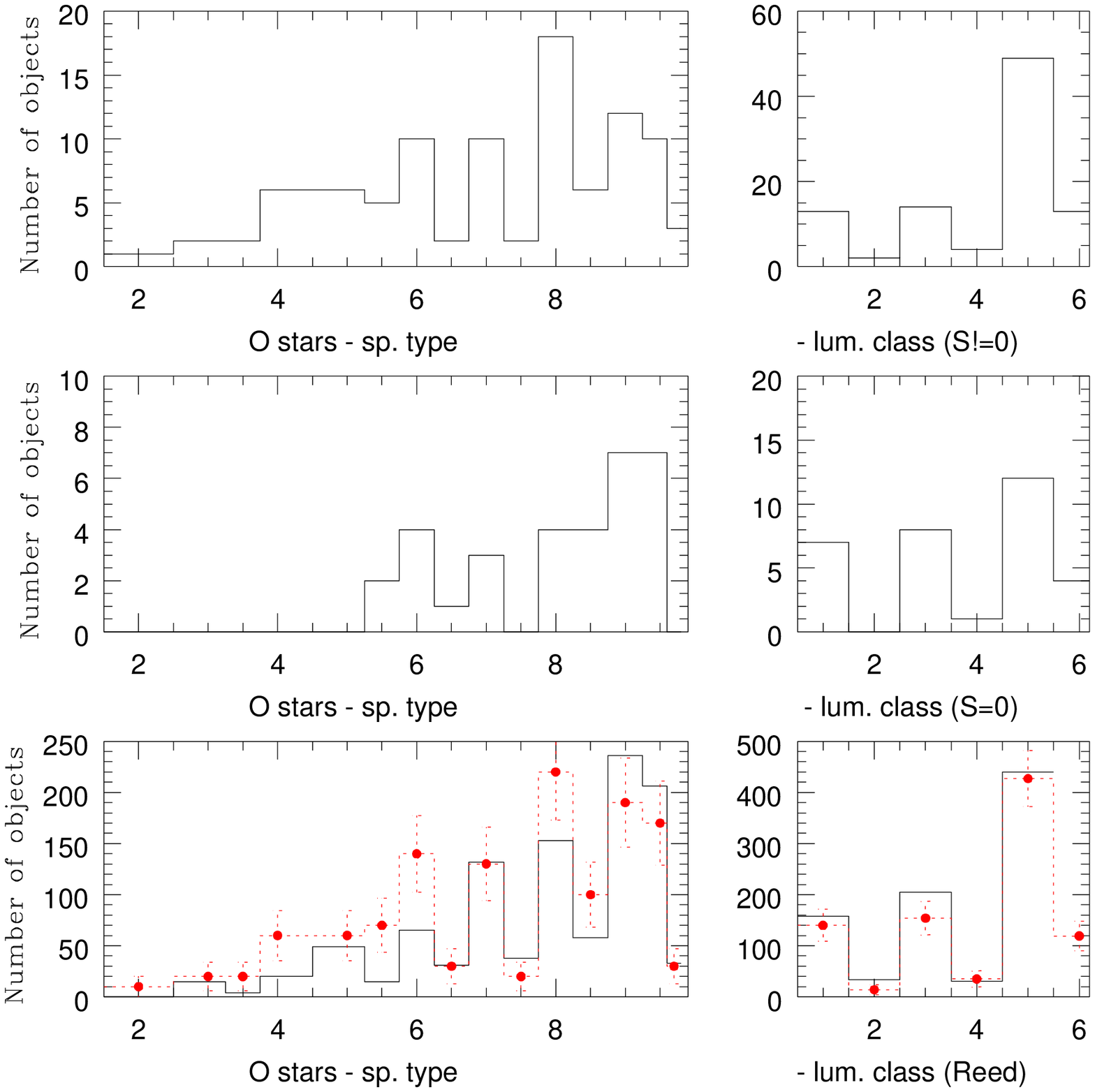}
   \includegraphics[width=9cm]{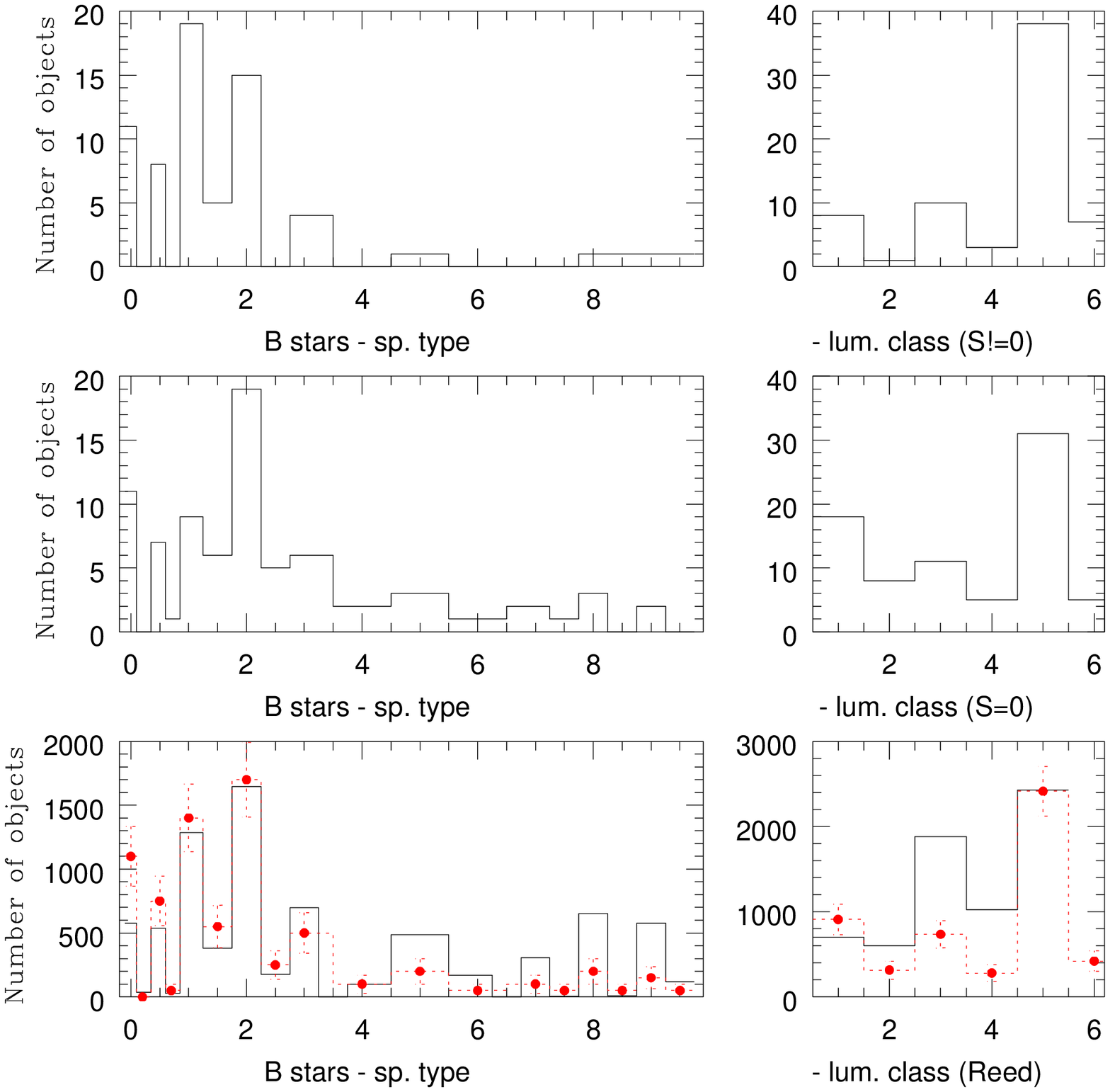}
   \caption{Distribution of the spectral types and luminosity classes of the O and B stars detected in X-rays (when detailed classification is available; for binaries, only the classification of the primary was considered). The middle panels correspond to X-ray sources with a good quality flag ($S$=0), while the upper panels show the results for ``potentially spurious'' sources ($S$!=0). The corresponding distributions for the entire Reed catalog are shown in the lowest panels, where the dotted red line indicates the scaled distribution of all detected objects (zero and non-zero quality flags, with Poissonian errors). Note that a luminosity class of 6 indicates stars with known spectral type but undefined luminosity class.}
              \label{repart}
    \end{figure*}

\setcounter{table}{1}
\begin{table}
\caption{Classification of the optical sources detected in the 2XMMi catalog. The number in parentheses yields the number of sources with available EPIC spectra.}
\label{tab:clas}
\centering
\begin{tabular}{l c c c}
\hline\hline
Quality flag & O stars & B stars & WR stars \\     
\hline                        
0 & 35 (16) & 91 (22) & 7 (6)\\ 
$\neq$0 & 98 (62) & 70 (28) & 9 (8)\\ 
\hline                                   
\end{tabular}
\end{table}

\begin{table}
\caption{Percentage of known binaries amongst the detected hot stars.}
\label{tab:bin}
\centering
\begin{tabular}{l c c c}
\hline\hline
Quality flag & O stars & B stars & WR stars \\     
\hline                        
0 & 17\% (6/35) & 5\% (5/91) & 57\% (4/7)\\ 
$\neq$0 & 22\% (22/98) & 4\% (3/70) & 89\% (8/9)\\ 
\hline                                   
\end{tabular}
\end{table}

\subsubsection{Distribution}

The repartition of the detected stars amongst the main spectral types is given in Table \ref{tab:clas} and shown in Fig. \ref{repart}. For sources with a good quality flag, the majority display a B spectral type, as one could have expected based on the incidence of these lower-mass objects compared to O and WR stars. When potentially spurious sources are included, the situation appears to change because the O stars now dominate. This is probably caused by the combination of two facts: (1) the O stars are brighter in X-rays, and (2) these ``spurious'' sources are found mostly in clusters, where only bright O-type stars are easily detected against the overall background produced by the combined X-ray emission from PMS stars, background objects, and diffuse emission within the cluster. 

Interestingly, it must be noted that, excluding WR systems, binarity does not seem to play an important role in the probability of X-ray detection (see Table \ref{tab:bin}: only $\sim$20\% of the detected O stars and $\sim$5\% of the detected B stars are known binaries). This result might first seem at odds with previous conclusions based on older data \citep[e.g., $Einstein$ observations in][]{chl89}. Although one cannot exclude the possibility that part of the binary population remains undiscovered (especially for the B stars), this bias cannot explain the new result since the older data were affected by similar uncertainties; besides, it would indeed be unexpected that \xmm\ targeted a specific area of the sky with a low number of binaries. More importantly, it must be recalled that the preferred detection of binaries was then explained by the presence of an additional X-ray emission produced by wind-wind collisions. However, it was shown that, in sensitive observations of an entire hot-star population, only a small fraction of the massive binaries display strong wind-wind collisions capable of emitting X-rays: only a few systems are thus truly overluminous in the 0.5--10\,keV range \citep{san06}. This is linked to the X-ray excess, when detected, being possibly dependent on the bolometric luminosity and the orbital parameters \citep{chl91,lin06}. Therefore, there may be two different causes of the apparent discrepancy between more recent observations and the first $Einstein$ results: the sample size (a few bright, peculiar objects compared to a larger sample of many different cases), as well as improvements in the knowledge of the physical parameters (e.g., with respect to stellar multiplicity, especially the detection of faint companions) and in the X-ray instrumentation (higher spatial/spectral resolution and higher sensitivity, including at lower energies than $Einstein$, i.e., where the wind-wind emission is less prominent than at higher energies).

Finally, the distribution of the spectral types and luminosity classes is shown in Fig. \ref{repart}. Amongst O-type stars, the more numerous O6--9.7 stars constitute most of our sample. This is also true for main-sequence systems, which are more numerous in general, this study being no exception. The O-star distribution well reflects that of the initial stellar catalog, as can be seen in the bottom panels of Fig. \ref{repart}. However, there is a lack of detections amongst late B-type stars, while earliest B-stars as well as giant B-stars appear preferentially detected. At first sight, the former results appear in phase with the current wind-shock model: one does expect the X-ray luminosity to drop (hence the detection rate) as the stellar wind weakens and then disappears. 

\begin{table*}
\caption{Number and percentage of variable X-ray counterparts. When comparing several exposures, the percentages noted for non-tested objects were calculated with respect to the entire sample, whereas those of the variable sources refer only to the tested subsample.}
\label{tab:var}
\centering
\begin{tabular}{c c| c c c | c c c c}
\hline\hline
Q. flag & Sp. Type & \multicolumn{3}{c|}{Within one exposure} & \multicolumn{3}{c}{Between exposures} &\\     
& & total & single & binaries & total & single & binaries & not tested\\     
\hline                        
0 & O  & 0 & 0 & 0 & 4 (25\%)& 3 (25\%) & 1 (25\%) & 19 (54\%)\\ 
0 & B  & 4 (4\%) & 4 (5\%) & 0 & 7 (32\%)& 7 (33\%)& 0 & 69 (76\%)\\ 
0 & WR & 0 & 0 & 0 & 2 (50\%)& 1 (50\%)& 1(50\%) & 3 (43\%)\\ 
\hline                                   
$\neq$0 & O  & 4 (4\%)& 3 (4\%)& 1 (5\%)& 44 (69\%)& 32 (67\%)& 12 (75\%)& 34 (35\%)\\ 
$\neq$0 & B  & 4 (6\%)& 4 (6\%)& 0 & 14 (67\%)& 12 (63\%)& 2 (100\%) & 70 (49\%)\\ 
$\neq$0 & WR  & 0 & 0 & 0 & 6 (86\%)& 0& 6 (100\%) & 2 (22\%)\\ 
\hline                                   
\end{tabular}
\end{table*}

\subsubsection{Short- and long-term variations}

Table \ref{tab:var} summarizes the results of the variability study.
The short-term variability was directly analyzed during the 2XMMi processing. The first variability flag quoted in Table 1 corresponds directly to the variability flag of the 2XMMi catalog, i.e., it is defined to be 1 if the source was found to be variable with a significance level of 0.001\% following a $\chi^2$ test performed on the time series of the individual exposures. For hot stars, as could be expected, very few objects vary within one exposure ($<$10\%). Three sources display a flare, which is quite typical of low-mass,  pre-main-sequence stars: HD37016 (B2.5V), HD33904 (B9III), and HD37479 ($\sigma$\,Ori\,E, B1/2V, whose flare decay has a rather long time constant). In addition, HD120991 (B2Ve) exhibits a clear increase of its count rate during the observation. For the other stars, there is no obvious flare, and the variations cannot easily be differentiated from those of the background signal. 

Long-term variability, for example $between$ exposures, can also occur, but the 2XMMi catalog does not check for its presence. A second variability flag was thus calculated for Table 1: it is set to be 2 if there are not enough exposures (0 or 1 observation for all combinations of filters/instruments), to 1 if a $\chi^2$ test detected variations in the count rate with a significance level of 1\%, and otherwise to 0. As is obvious from Table \ref{tab:var}, this type of variability is far more common in hot stars, with few differences between single and binary OB stars. Except for the putative presence of colliding-wind binaries or magnetic wind confinement, the cause of these variations remains unknown.

\subsection{Spectral fitting for the X-ray bright objects}

About half of the detected hot stars (59\% of the O stars, 31\% of the B objects, and 88\% of the WRs) have enough counts within a single exposure to have their spectra automatically extracted by the 2XMMi processing (Table \ref{tab:clas}). Many of these were observed several times: the individual spectra were merged, instrument by instrument, only if the source was non-variable. In total, 332 spectra of OB stars were finally fitted. They were individually fitted, in the 0.3--10.\,keV energy range, within Xspec v11.2.0 with absorbed multi-temperature thermal plasma models of the type $wabs(ISM)\times wabs\times (\sum mekal_i)$ with i equal to 3 at most and solar abundances assumed. The first absorbing column was fixed to the ISM column derived above (see Sect. 2) if known or to 0 otherwise. The second absorbing column allows for additional, local absorption. A word of caution should be added: the soft and hard thermal components might not be formed at the same location inside the wind and might thus be affected by different local absorptions. However, we restrict ourselves to one common local absorbing column because, for the majority of the stars investigated in this paper, the quality of the data does not justify the use of a model of the type $wabs(ISM)\times (\sum wabs_i \times mekal_i)$. The number of optically-thin plasma components is the minimum number of single-temperature components needed to provide a good fit ($\chi^2\sim1$ or $<2$) to the considered data. Generally, the 1T-models are used only when there are too few counts, i.e., only the main component is distinguishable. When the signal-to-noise ratio is reasonably high, the sum of two thermal components is requested to provide a good fit. 

Table 5 (available electronically at the CDS) provides the results of these fits. The first column indicates the Reed number of the objects, the second column their usual name, and the third column the interstellar absorbing column ($N^{\rm ISM}_{\rm H}$, in cm$^{-2}$). The fourth and fifth columns provide the value of $\chi^2$ and the number of degrees of freedom. The next four columns list the absorbed fluxes (in erg s$^{-1}$ cm$^{-2}$) in the 0.5-10.\,keV,  0.5-1.\,keV, 1.-2.5\,keV, and 2.5-10.\,keV energy bands. They are followed by 4 similar columns providing the unabsorbed fluxes (i.e., the X-ray fluxes dereddened by the insterstellar component, if known, or otherwise by the fitted absorption column). The additional absorption is given in Col. 14 ($N_{\rm H}$, in 10$^{22}$ cm$^{-2}$), followed in the next two columns by the lower and upper limits of the 90\% confidence interval. The temperature (k$T$ in keV) and normalization factors ($norm$ in cm$^{-5}$) of each $mekal$ component in turn (in order of increasing temperatures) are then given, together with their 90\% confidence interval. Finally, the last column provides additional remarks, such as the observation identifiers (if several exposures were available) or, in a few cases, any modification of the general fitting scheme (e.g., non-solar abundances). Undetermined values are again quoted as ``NULL''.  Note that a comparison was made between our results and published analyses of some hot stars' spectra (Of?p stars, \citealt{naz04,naz07,naz08b}; $\theta$\,Car, \citealt{naz08c}; NGC6231, \citealt{san06}; HD168112 \citealt{deb04}; Cyg OB2 \#8A, \citealt{deb06}; Carina OB1, \citealt{ant08}) and a good agreement was found. It must also be emphasized that these fits provide an adequate idea of the flux distribution for the spectral resolution and signal-to-noise ratio considered, but they should not be taken at face value (especially in the case of non-solar abundances), since the true properties of the hot plasma can differ (e.g., continuous distribution of thermal components versus a few discrete temperatures).

\setcounter{table}{5}
   \begin{figure*}
   \centering
   \includegraphics[width=4.5cm, bb=200 160 575 700, clip]{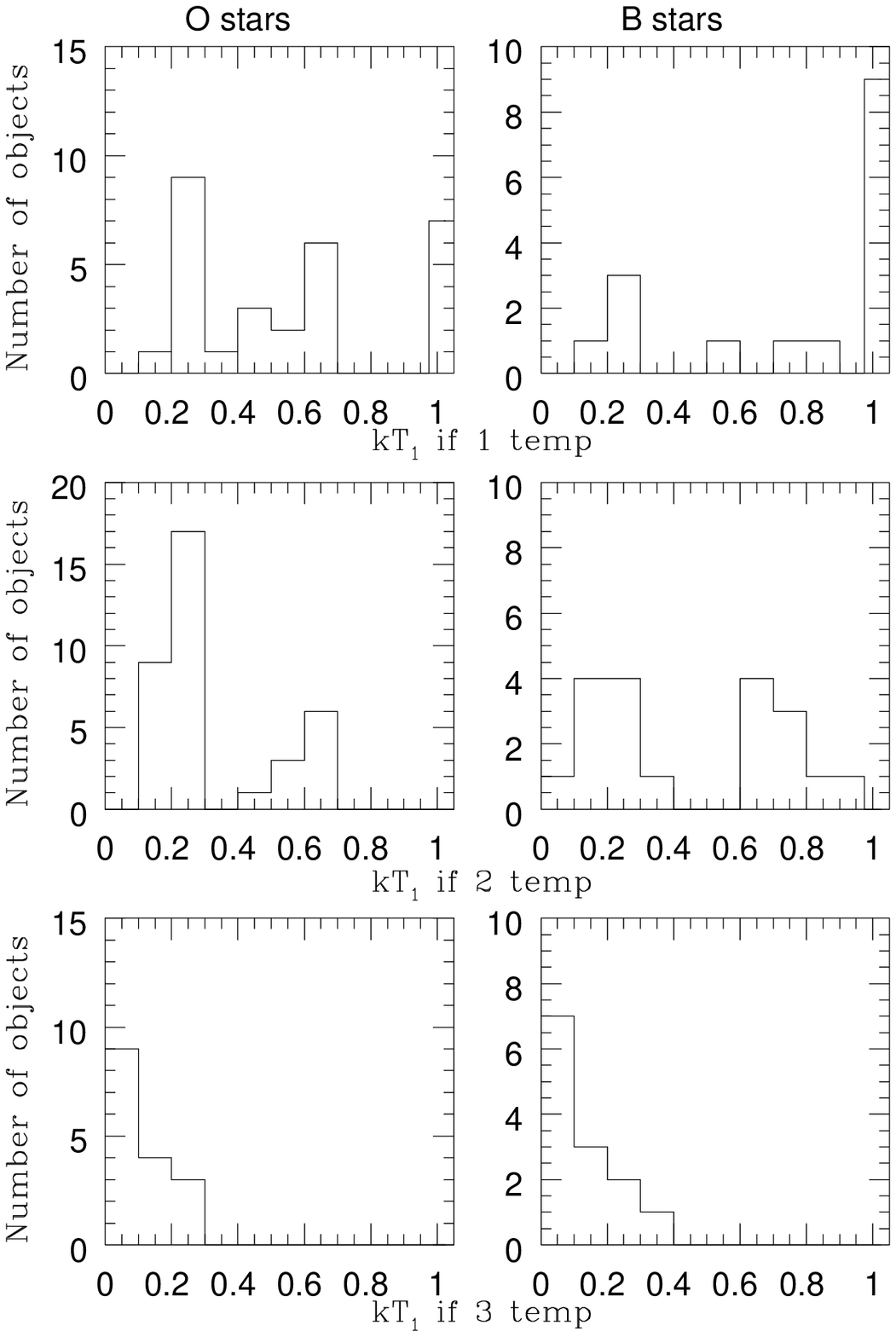}
   \includegraphics[width=4.5cm, bb=200 160 575 700, clip]{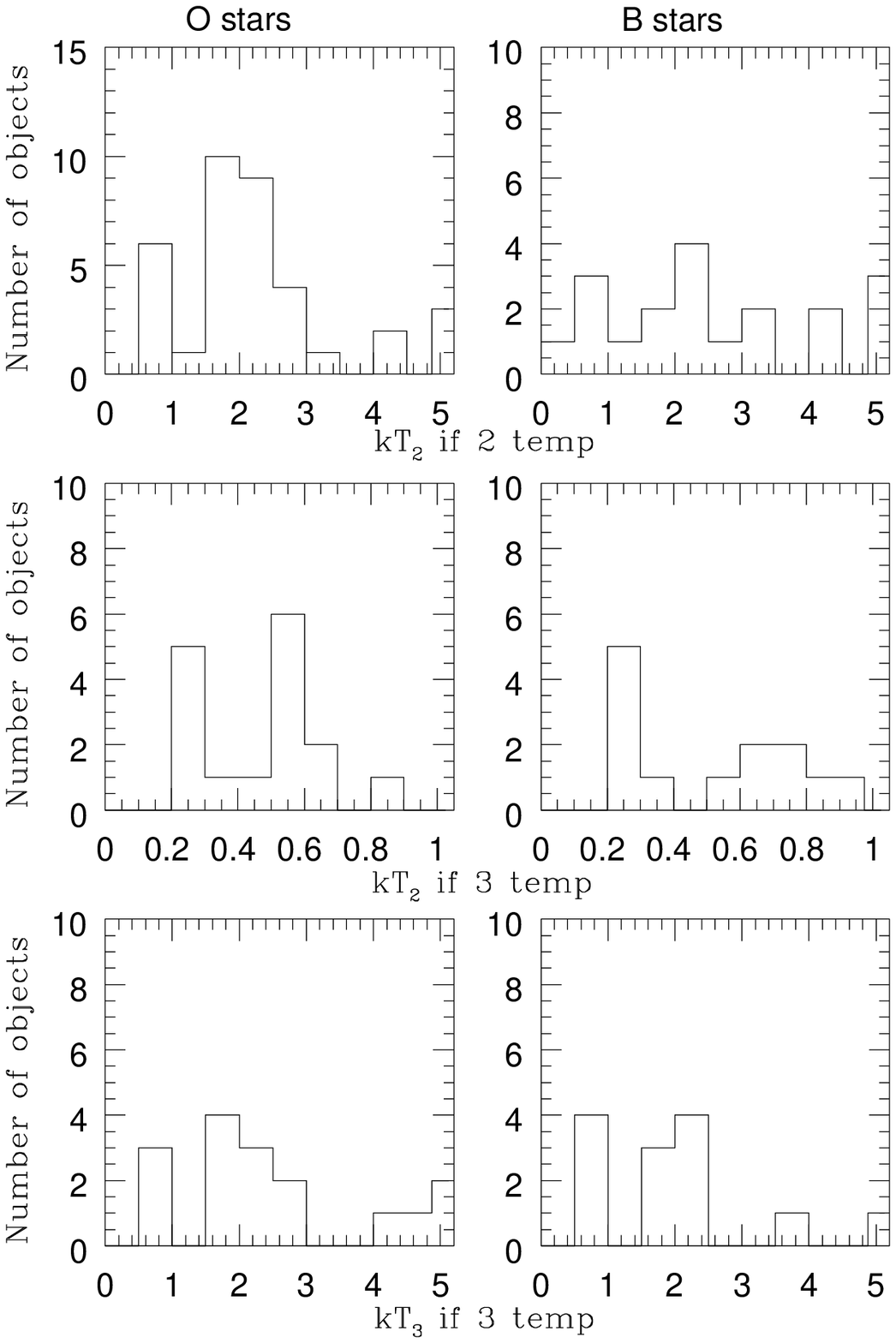}
   \includegraphics[width=4.5cm, bb=200 160 575 700, clip]{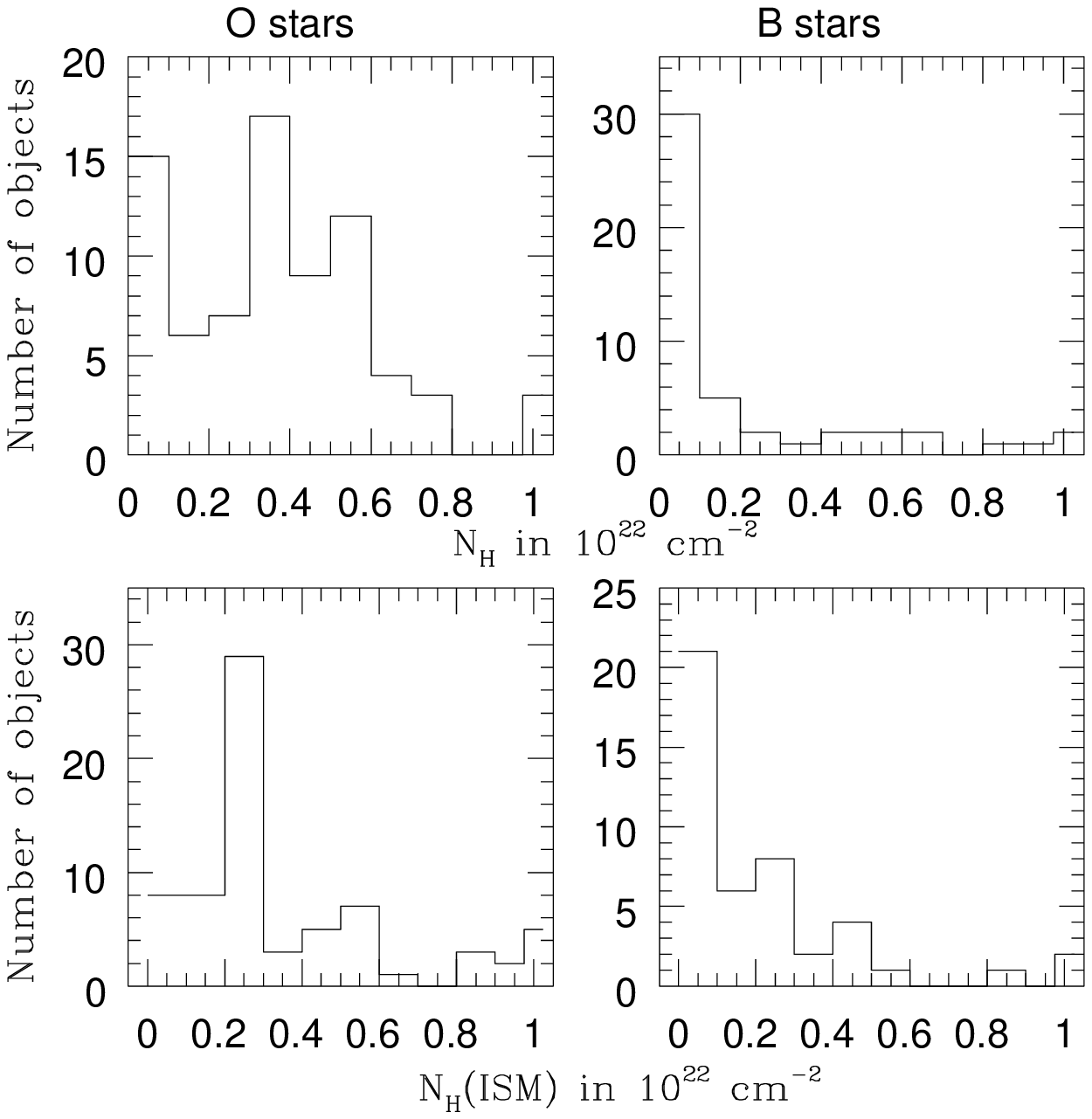}
   \includegraphics[width=4.5cm, bb=200 160 575 700, clip]{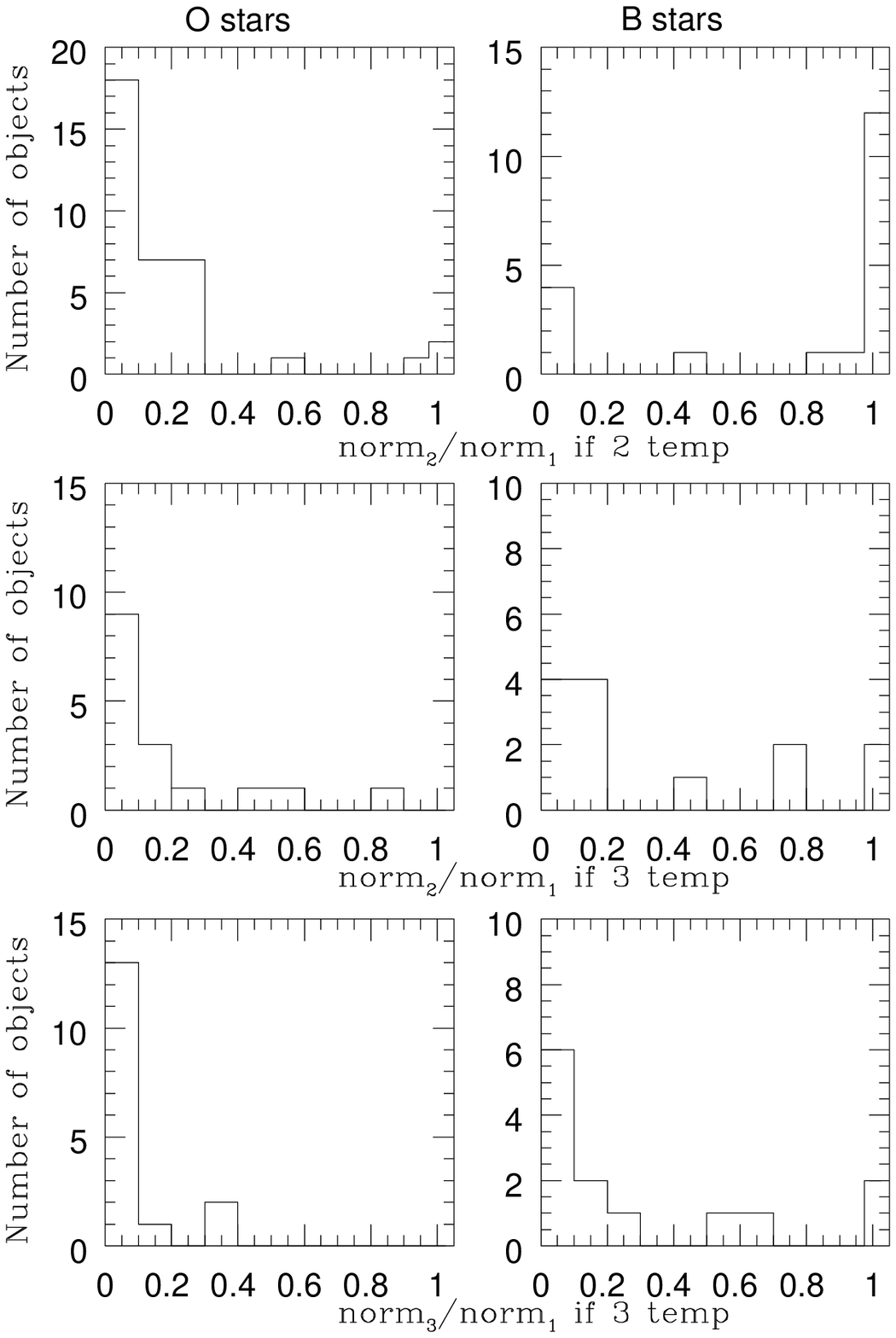}
   \caption{Distribution of the spectral properties (temperatures k$T_i$, absorptions $N^{\rm ISM}_{\rm H}$ and $N_{\rm H}$, ratio of normalization factors norm$_i$/norm$_1$) for the whole dataset, the O-type stars only and the B-type objects only. Note that, when several fits were made, an average value was used in the computation of the histogram and that bins at the rightmost limits of the plots correspond to all values larger than this limit. }
              \label{fits}
    \end{figure*}

Figure \ref{fits} graphically illustrates the results. In terms of the absorption, the interstellar components appear larger for O stars than for B stars, as expected (O stars being brighter and hence detectable from further away). In addition, a significant, additional absorption is needed when fitting O-star spectra. Since its distribution does not correlate with the interstellar absorption histogram, this effect should be considered to be significant. Concerning temperatures, different cases corresponding to different number of $mekal$ components should be envisaged. If only one thermal component was fitted, its temperature is usually quite low, 0.2 or 0.6\,keV, for O stars, while the distribution is much flatter for B stars, temperatures in half of the cases being below 1 keV and in the other half above. If two thermal components are fitted, the first temperature is generally quite soft, with a frequent ambiguity between 0.2 and 0.6\,keV (for previous reports, see e.g., the case of HD148937 in \citealt{naz08b}). The second temperature is about 2\,keV for O stars and B stars, but has a far larger dispersion than in the latter case. If three thermal components were fitted, the first temperature was always very low (often at the lowest limit of the permitted range within Xspec, 0.08\,keV), the second temperature again is 0.2 or 0.6\,keV, while the third one is generally 2\,keV. The normalization factors of the additional components usually have lower values than that of the first component, except in the case of 2-temperature fits to B stars, where the harder component clearly dominates.

\subsubsection{X-ray fluxes}

Spectral modelling directly provides the X-ray fluxes of each fitted object (see Table 5), which are simply reproduced in Table 1. In the case of a varying source observed several times, the X-ray flux reported in Table 1 is the average flux.

However, the objects in the fainter half of the sample do not have spectra, and any estimate of their flux must rely only on their count rates. To convert count rates into fluxes, the average/the most probable 2T-model was used. For O-type stars, it has a local absorbing column of $\sim$4$\times10^{21}$\,cm$^{-2}$ and temperatures of 0.2 and 2\,keV, which is represented well by the spectrum of HD168112 (ALS 4912, second observation): the latter was therefore used as a reference. For B-type stars, the average properties are a zero local absorbing column and temperatures of 0.2 or 0.6 and 2\,keV, and a good reference in this case is HD37040 (ALS 14653). The spectral properties of these two reference stars ($N_{\rm H}$,k$T_{1,2}$, flux ratio at 1 keV of 0.158 and 1.46, respectively) were used as input for PIMMs\footnote{http://heasarc.gsfc.nasa.gov/Tools/w3pimms\_pro.html}. The unabsorbed fluxes (i.e., dereddened only for the interstellar component) were calculated for a unit count rate, one of the two cameras, one of the three filters, and a range of interstellar absorbing columns (from 0 to 2.75 $\times10^{21}$\,cm$^{-2}$). The resulting conversion factor had to be corrected by a factor of 0.70 because PIMMs considers count rates in a region of 15'' radius while the 2XMMi catalog provides the full count rate \citep{wat09}\footnote{In addition, PIMMs considers only counts with a zero PATTERN keyword, while the 2XMMi counts are for PATTERN keywords ranging from 0 to 12 (MOS) or 0 to 4 (pn). However, as most of the recorded counts have a zero PATTERN, this uncertainty is considered to be negligible compared to other sources of error.}. For each star, the X-ray flux was calculated by taking into account its own interstellar absorption (see Sect. 2), no value was thus quoted if this absorption is unknown. The results were checked against the bright stars and good agreement was found - in 80\% of the cases, the flux derived from the count rates is within a factor of 2 of the flux derived from the spectral fits. The derived unabsorbed X-ray fluxes for the faintest stars can be found in Table 1. 

\begin{figure*}
   \centering
   \includegraphics[width=8.5cm]{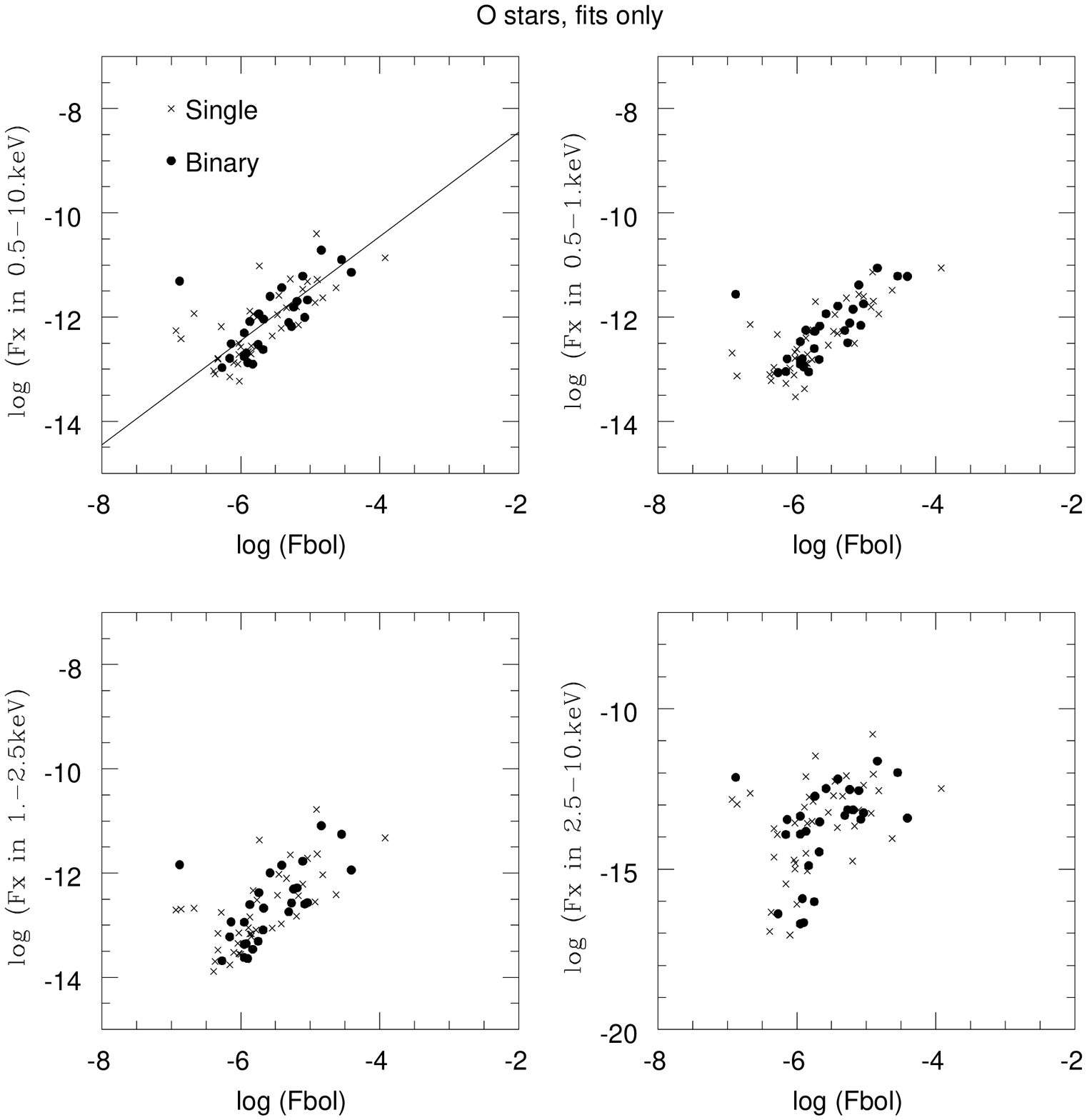}
   \includegraphics[width=8.5cm]{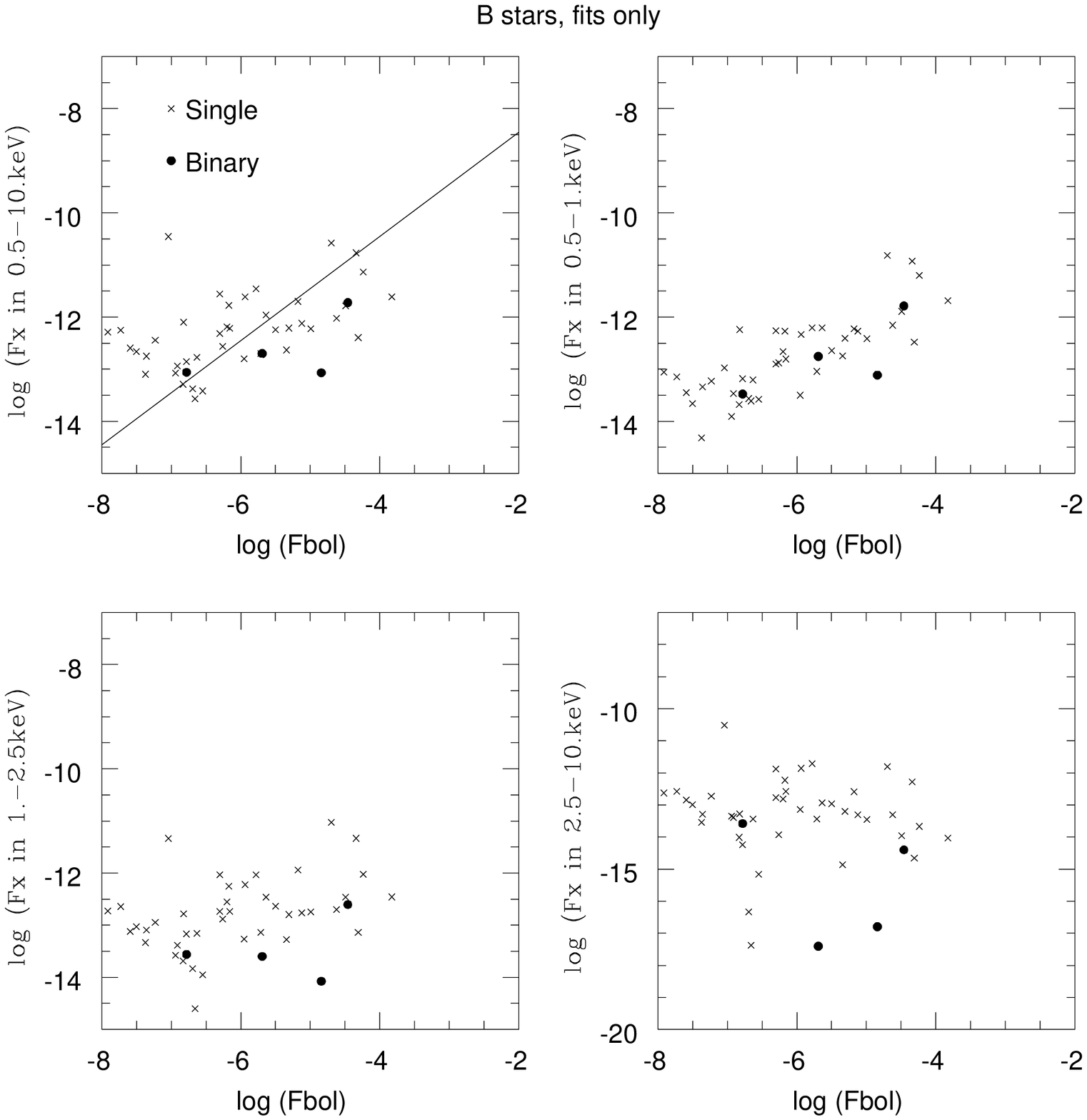}
   \includegraphics[width=8.5cm]{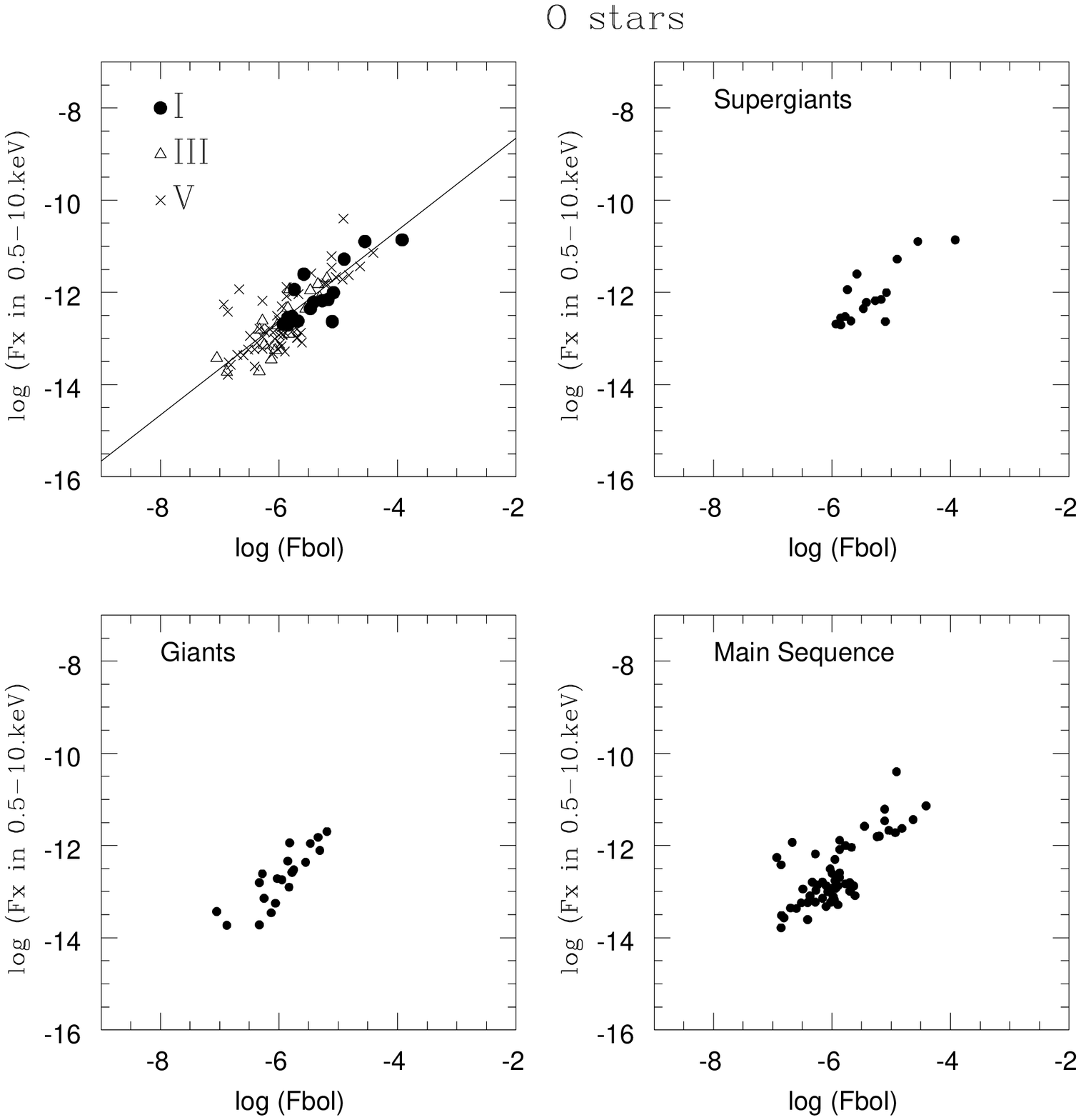}
   \includegraphics[width=8.5cm]{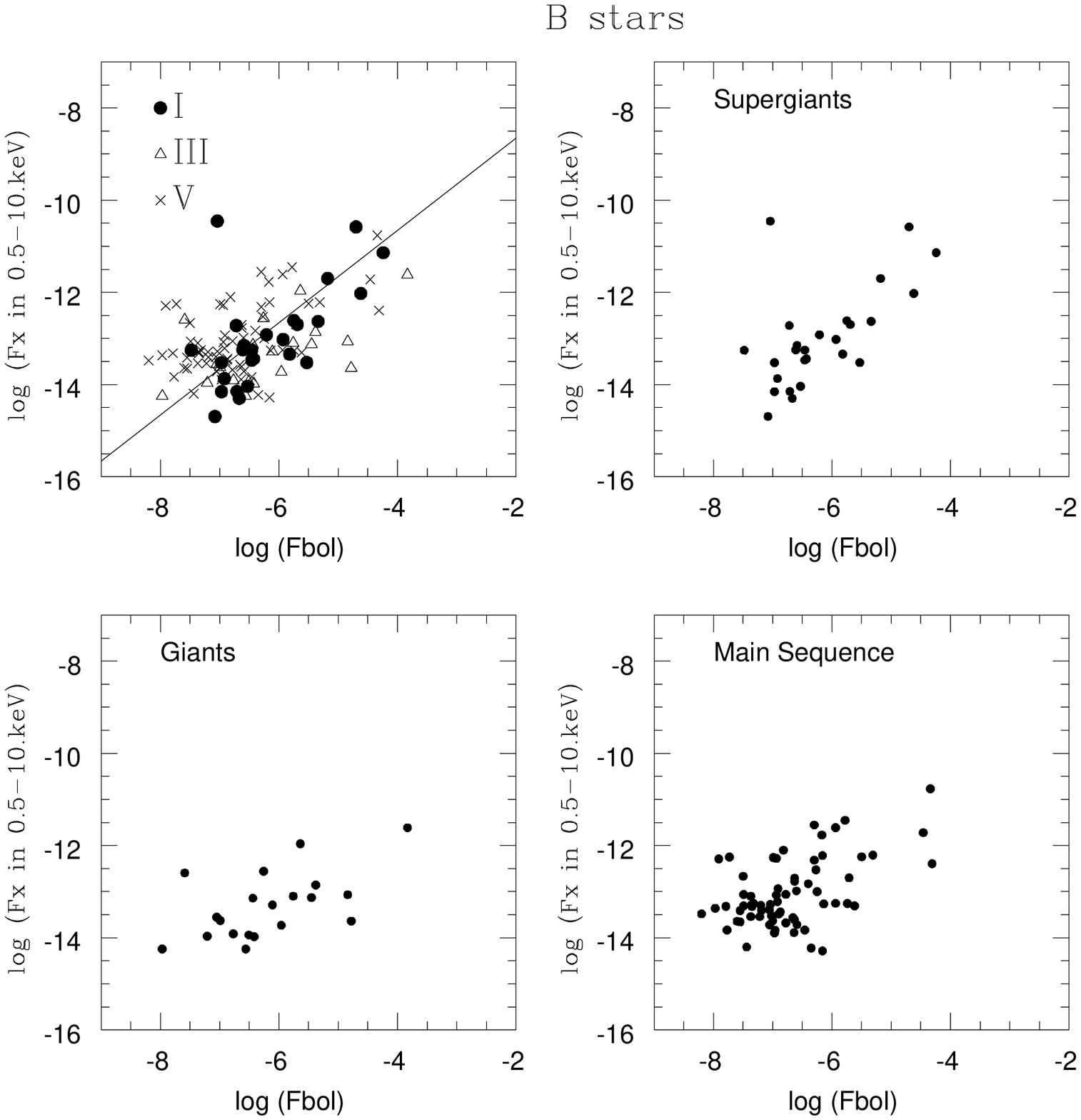}
   \includegraphics[width=8.5cm,bb=18 432 591 717,clip]{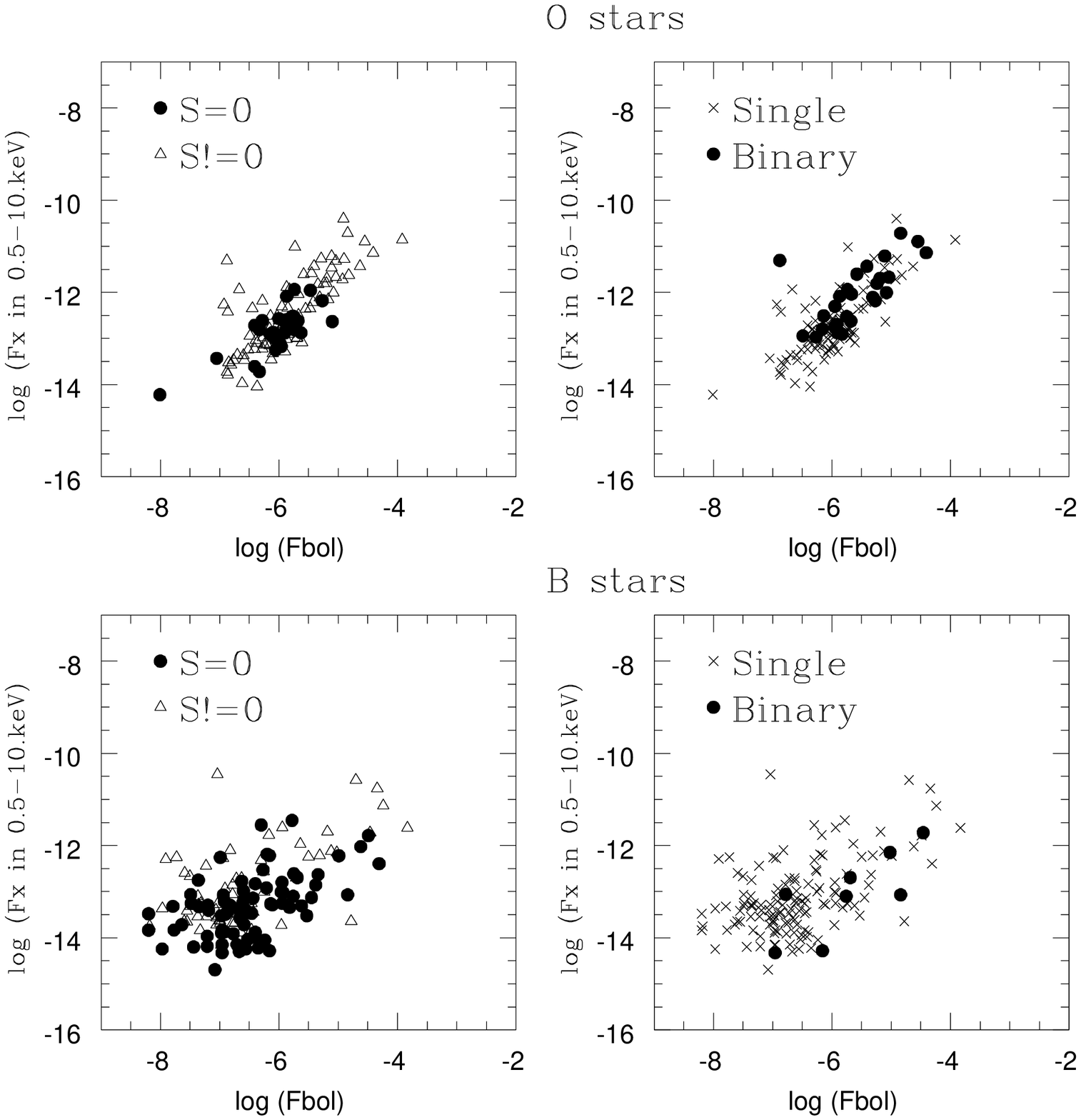}
   \includegraphics[width=8.5cm,bb=18 153 591 432,clip]{flux_final_qualitybinary.ps}
   \caption{Unabsorbed X-ray flux as a function of the bolometric flux. Lines indicate the average $\log f_{\rm X}/f_{\rm BOL}$ ratio from Table \ref{tab:fxfbol}. In the case of binaries, the luminosity class of the primary was used as a reference in the middle panels.}
              \label{fxfbol}
    \end{figure*}

Once the fluxes are known, the $\log f_{\rm X}/f_{\rm BOL}$ ratios, equivalent to the $\log L_{\rm X}/L_{\rm BOL}$ ratio, were derived. They are shown in the last column of Table 1. Note that, for the five pairs of stars sharing the same 2XMMi counterpart, the ratio is calculated for the total bolometric flux. The unweighted average $\log f_{\rm X}/f_{\rm BOL}$ ratios (for all stars but the pairs) were then calculated, together with their dispersion. The results are listed in Table \ref{tab:fxfbol} and shown in Fig. \ref{fxfbol}\footnote{It might at first appear surprising that the detected O and B stars display similar bolometric fluxes. It must however be kept in mind that the $\log L_{\rm X}/L_{\rm BOL}$ ratios for these stars are rather similar: if the X-ray sources constitute a flux-limited sample, the bolometric fluxes of the detected objects will also appear quite similar. Indeed, this conclusion does not apply to bolometric $luminosities$: as shown in Fig. \ref{fits}, the O stars of our sample are more absorbed (hence intrinsically much brighter after correcting for the ISM absorption), and much more distant, than the B stars. }

\begin{table}
\caption{Average $\log f^{unabs}_{\rm X}/f_{\rm BOL}$ ratios, with their dispersions.}
\label{tab:fxfbol}
\centering
\begin{tabular}{l c c }
\hline\hline
 & O stars & B stars \\     
\hline                        
\multicolumn{3}{l}{From fits only}\\
\vspace*{-0.2cm}\\
0.5--10\,keV & $-6.45\pm0.51$ & $-6.27\pm0.98$ \\ 
\vspace*{-0.2cm}\\
0.5--1.\,keV & $-6.69\pm0.45$ & $-6.73\pm0.69$ \\ 
1.--2.5\,keV & $-7.02\pm0.58$ & $-6.82\pm1.07$ \\ 
2.5--10\,keV & $-8.03\pm1.32$ & $-7.48\pm1.89$ \\ 
\vspace*{-0.2cm}\\
\multicolumn{3}{l}{From fits and conversion of the count rates (0.5--10\,keV band)}\\
\vspace*{-0.2cm}\\
all     & $-6.66\pm0.50$ & $-6.59\pm0.83$ \\ 
\vspace*{-0.2cm}\\
$S$=0   & $-6.80\pm0.35$ & $-6.81\pm0.74$ \\ 
$S\neq$0& $-6.61\pm0.54$ & $-6.34\pm0.87$ \\ 
\vspace*{-0.2cm}\\
binaries& $-6.47\pm0.52$ & $-7.34\pm0.62$ \\ 
singles & $-6.72\pm0.49$ & $-6.55\pm0.82$ \\ 
\vspace*{-0.2cm}\\
supergiants  & $-6.75\pm0.36$ & $-6.82\pm0.88$ \\ 
giants       & $-6.74\pm0.33$ & $-7.13\pm0.85$ \\ 
main sequence& $-6.67\pm0.49$ & $-6.33\pm0.77$ \\ 
\vspace*{-0.2cm}\\
Carina  & $-6.54\pm0.56$ &\\ 
NGC6231 & $-6.71\pm0.22$ &\\ 
NGC6604 & $-6.48\pm0.48$ &\\ 
CygOB2  & $-6.97\pm0.40$ &\\ 
\hline                                   
\end{tabular}
\end{table}

\section{Detections of hot stars in the \xmm\ Slew Survey}

Between pointed observations, \xmm\ slews between different sky areas. During many of these slews, the telescopes remain open and thus perform a limited survey. The analysis of the EPIC-pn data from 218 slews detected more than 2600 X-ray sources \citep{sax08}. They constitute the XMMSL1 catalogue, which can be queried using Vizier\footnote{http://vizier.u-strasbg.fr/viz-bin/VizieR?-source=J/A+A/480/611}. 

As for the 2XMMi catalog, the results of the slew survey were cross-correlated with the preferred names and coordinates extracted from the Reed catalog, again for a correlation radius of 5'', yielding 13 detections.  Extended X-ray sources and sources already belonging to the 2XMMi catalog were discarded from the list, as well as wrong identifications (i.e., a hot star in the vicinity of an X-ray source, but where the hot star is not the ``official'' counterpart listed in the slew survey catalog). In addition, counterparts were automatically identified for slightly more than half of the X-ray sources of the XMMSL1: their nature was checked using Simbad, which inferred that they were 34 X-ray emitters associated with hot stars, sometimes more distant than 5''. Of these, those stars not belonging to the Reed catalog or those already found in the 2XMMi catalog were discarded. This two-step trimming process finally resulted in the detection of 16 additional X-ray sources, presented in Table \ref{tab:slew}, where the total count rates are given for the pn instrument with the medium filter and in the 0.2--12.\,keV band, and the DL column refers to the individual detection likelihoods. 

As for the 2XMMi data, the optical properties of the star (e.g., interstellar reddening, bolometric flux) were derived from their known characteristics (e.g., spectral type, magnitudes). The count rates were then converted into unabsorbed fluxes in the 0.5--10.\,keV band, in a similar way to the method described in Sect. 3.2.1, to derive the $L_{\rm X}-L_{\rm BOL}$ ratio. 

\begin{sidewaystable*}
\caption{Additional X-ray detections of hot stars using the XMM slew survey. }
\label{tab:slew}
\centering
\begin{tabular}{l c l c c c c c l c c c c c }
\hline\hline
Reed name & ALS \# & Usual name & Sp. Type & Bin.?& V & B-V & d& XMMSL1 name & DL & Count rate & $\log(f_{\rm BOL})$& $f_{\rm X}^{unabs}$ & $\log(f_{\rm X}/f_{\rm BOL})$\\
& & & & & & & ('') & & & ct s$^{-1}$ & & erg cm$^{-2}$ s$^{-1}$ \\
\hline                        
HR 3129        &14910 &HD 65818        &B1V+B3:    &Y&  4.51&$-$0.18&  8.2&J075815.2$-$491444&  9.7 & 0.55$\pm$0.31 &$-$5.23 & 1.01e-12& $-$6.77\\   
LS III +44  43 &11848 &BD +43 3913     &B1.5V      & &  8.91&  0.54 &  6.4&J212501.9+442706  & 15.8 & 0.57$\pm$0.20 &$-$6.18 & 3.68e-12& $-$5.25\\   
LSE 99         &18922 &CD$-$35 9665    &~O+        & & 12.70&       &  4.3& J143740.0$-$36132&      & 1.85$\pm$0.58 & & & \\ 
LS  4635       & 4635 &HD 165052       &O6V((f))   &Y&  6.87&  0.09 &  3.2&J180510.3$-$242354&      &               &$-$5.47 & & \\        
LS III +43  19 &11807 &HD 203064       &O7.5III    &Y&  5.04&$-$0.06&  5.0&J211827.7+435646  & 10.8 & 0.65$\pm$0.25 &$-$5.06 & 1.68e-12& $-$6.72\\  
HR 3860        &16348 &HD 83979        &B5V        & &  5.07&$-$0.14&  1.0&J093353.1$-$805629&      & 0.70$\pm$0.38 &$-$6.01 & 1.11e-12& $-$5.94\\  
HD 157832      &17479 &HD 157832       &B3Ve       & &  6.66&  0.00 &  7.7&J172754.0$-$470132& 20.1 & 1.07$\pm$0.32 &$-$6.25 & 2.57e-12& $-$5.34\\
LS   V +35   1 & 7841 &HD 24912        &O7.5III(f) & &  4.04&  0.00 &  7.0&J035857.3+354731  & 11.4 & 1.55$\pm$0.69 &$-$4.59 & 4.58e-12& $-$6.75\\
HD 212571      &14728 &HD 212571       &B1III/IVe  &Y&  4.80&$-$0.18& 16.4&J222517.0+012223  & 18.6 & 2.40$\pm$0.73 &$-$5.46 & 4.41e-12& $-$5.90\\
HR 6510        &15050 &HD 158427       &B3V        & &  2.83&$-$0.11&  5.8&J173149.8$-$495233& 15.4 & 0.84$\pm$0.31 &$-$4.85 & 1.59e-12& $-$6.95\\
HD 133242      &17476 &HD 133242       &B5V        & &  4.59&$-$0.14&  3.5&J150507.4$-$470303&      & 1.04$\pm$0.40 &$-$5.82 & 1.66e-12& $-$5.96\\
Hbg 1339       &14476 &HD 155806       &O7.5IIIe   & &  5.62&$-$0.04&  8.6&J171519.4$-$333246& 17.2 & 0.72$\pm$0.25 &$-$5.27 & 1.95e-12& $-$6.44\\
LS  4148       & 4148 &WR93            &WC7+O7/9   &Y&      &       & 15.7&J172507.6$-$341110& 14.5 & 0.53$\pm$0.19 & & & \\
MCW 441        &14798 &HD46328         &B1III      & &  4.33&$-$0.24&  6.0&J063150.9$-$232509& 16.6 & 1.40$\pm$0.42 &$-$5.34 & 2.16e-12& $-$6.33\\
HR 3055        &14901 &HD63922         &B0III      & &  4.10&$-$0.20&  4.7&J074913.9$-$462227&      & 0.96$\pm$0.27 &$-$4.98 & 1.82e-12& $-$6.76\\
LS  4142       & 4142 &HD319718        &O3If*      & & 10.43& 1.45  &  4.6&J172443.5$-$341201&      & 0.69$\pm$0.22 &$-$5.11 & 8.10e-12& $-$5.98\\
\hline                                   
\end{tabular}
\end{sidewaystable*}

\section{Summary and discussion: the X-ray properties of OB stars}

Since the slew survey provided only a few additional sources with poorly constrained X-ray properties, the following conclusions rely mostly on the detailed analysis based on the 2XMMi catalog.

Since the \xmm\ observations cover only 1\% of the sky, a first question arise about the possible detection biases. For the detected O stars, the distribution of their spectral types and luminosity classes is fully comparable to that of the full hot star catalog from \citet{ree03}, so that no bias whatsoever is expected. This is not the case for the detected B stars, where a lack of late-type objects, as well as an excess of early-type stars and giant objects, are found. Although the consequence of selecting specific celestial regions should not be disregarded, it is probable that this effect is real: since the very first X-ray detection of hot stars, it has been shown several times that early-B stars are brighter, hence more easily detected, than late-type objects. 

Variability studies demonstrate that short-term changes are quite rare. When they exist, they generally appear flare-like, as found for PMS stars. It is as yet unclear whether the emission originates in a PMS object (true companions or line-of-sight coincidences), or if a magnetic phenomenon similar to that operating in PMS stars is at work in those objects. Stochastic variations such as those expected from collisions between wind-shells or large clumps are not seen. Long-term variations appear to be far more common in hot stars since they affect about half of the cases. They could well be caused by e.g., wind-wind collisions in unknown binaries, but will require an additional, in-depth analysis before a definite cause can be identified.

The spectral fits have enlighted the differences between O and B stars. The X-ray spectrum of O-type objects requires an absorption in addition to the interstellar component. This is not an effect of the uncertainties in the latter, which is quite well constrained since the intrinsic colour of O stars is quite similar regardless of subtype or luminosity class (although it is true that the $N_H-E(B-V)$ plots display some scatter; \citealt{boh78}). It thus appears intrinsic to these stars, most probably related to the presence of their stellar winds. Indeed, B-type stars, whose winds are much weaker, need not have any additional absorption. Another difference lies in the temperatures needed to fit  the plasma emission. On average, O-type stars exhibit a soft emission, well fitted by a thermal component with a temperature of 0.2 or 0.6\,keV, with possibly a faint hard component (typically at 2\,keV). On the other hand, the spectra of B-type stars are harder, well fitted by hot thermal components (either a single one at about 1\,keV, or the sum of a faint warm one with a temperature of 0.2--0.6\,keV and a brighter hot component with a temperature of 2\,keV). These hot components are not expected in the usual wind-shock model, and require the presence of either a PMS star along the same line-of-sight (physical companion or not) or of exotic magnetic phenomena. Analyses of multiwavelength follow-up observations would be required to settle this question.

Finally, the $L_{\rm X}-L_{\rm BOL}$ ratio was also investigated. Unsurprisingly, the dispersion is much lower for O stars than for B stars, and for the soft or medium energy bands than for the hard band (see e.g., similar results for specific clusters in \citealt{san06,ant08}). The relation for the B stars also appears shallower and might have a more complicated parametrization (e.g., $\log(L_{\rm X})=a\times \log(L_{\rm BOL})+b$ rather than $\log(L_{\rm X}/L_{\rm BOL})=b$) but this cannot be tested here because of a lack of accurate distances. For O-type stars, the dispersion in the $L_{\rm X}-L_{\rm BOL}$ relation is about 0.35--0.5 dex, or a factor of 2--3, a value similar to that found in the RASS analysis \citep{ber97}. Note that this dispersion is not mostly caused by uncertainties in the fluxes, since the fluxes are derived from precise fits, and those calculated from a general conversion of the count rates are similar in value. It thus seems that the dispersion measured by Bergh\"ofer et al. is real, and the much tighter correlation found for some clusters (e.g., NGC6231, \citealt{san06} and Table \ref{tab:fxfbol}) needs to be explained by reasons other than a different treatment of the data. Although the dispersions are somewhat large and the samples quite small, the  $L_{\rm X}-L_{\rm BOL}$ ratios found for different clusters indeed appear to differ marginally. 

   \begin{figure}
   \centering
   \includegraphics[width=8cm]{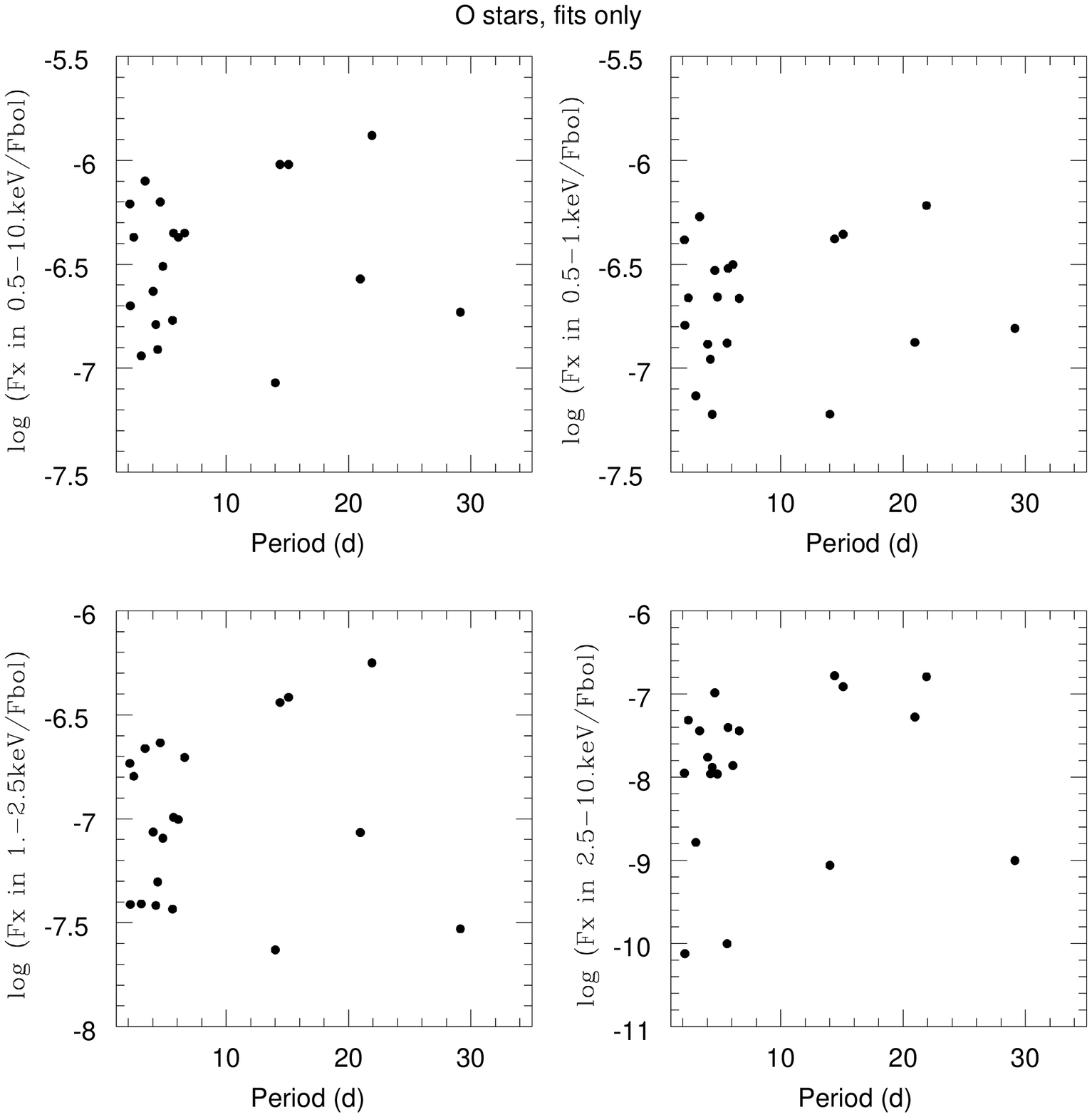}
   \includegraphics[width=8cm]{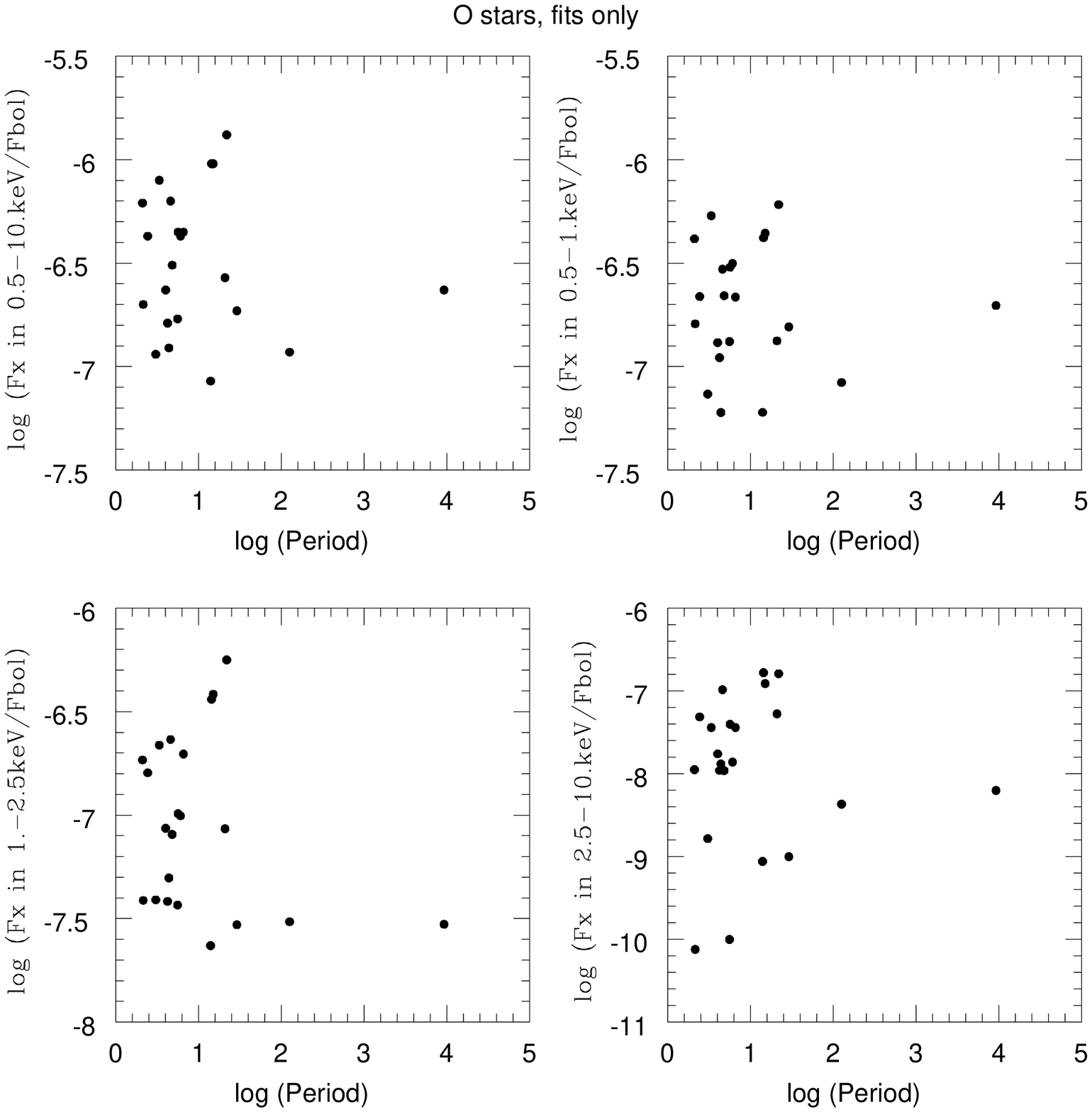}
   \caption{For the O+OB binaries with spectral information and defined period, $\log f_{\rm X}/f_{\rm BOL}$ ratio from Table \ref{tab:fxfbol}. }
              \label{overlumbin}
    \end{figure}

It must also be underlined that no significant difference was found between the $L_{\rm X}-L_{\rm BOL}$ ratios of objects with various quality flags, binary status, or luminosity classes, although we are only just below the 1$\sigma$ limit in the case of B-type stars. As discussed above (Sect. 3.1.1), binarity now appears less important than advocated in the past. One possible explanation for the weak impact of wind-wind interactions could be radiative inhibition, where the radiation of one component decelerates the wind originating in its companion \citep{gay97,ant04}. It should have more impact on short-period systems than on long-period binaries. Figure \ref{overlumbin} shows the variation in the \loglxlb\ ratio with the binary period, when the latter is known precisely (this is the case for 23 systems out of the 27 O+OB binaries having spectral information). For short-period systems (i.e., $P<35$\,d), there are hints of an overall shallow trend of increasing $\log f_{\rm X}/f_{\rm BOL}$ towards longer periods. Without a detailed modelling of each individual system, it is however difficult to assess whether the main cause of this trend is the radiative inhibition phenomenon or the expected $L_{\rm X}\propto\dot M v^2$ variation for radiative systems \citep{ste92}. It has to be emphasized that the scatter in the points of the $\log f_{\rm X}/f_{\rm BOL}$-period diagram is much larger than the mild ``trend" mentioned above. For long-period systems (i.e., $P>35$\,d), the average $\log f_{\rm X}/f_{\rm BOL}$ appears smaller than for short-period binaries, which might reflect the well-known $L_{\rm X}\propto\dot D^{-1}$ variation for adiabatic systems \citep{ste92}, but the lower limit to the $\log f_{\rm X}/f_{\rm BOL}$ ratio appears to increase towards longer periods, which might be related to the smaller impact of the radiative inhibition. However, the small number of systems in this range of periods (only 2!) prevents me from drawing solid conclusions. 


\section{Conclusions}

This paper presents the results of the first global survey of hot stars with the highly sensitive X-ray observatory \xmm. It relies on the 2XMMi and XMMSL1 catalogs. About 330 stars were detected, representing a sample comparable in size to the RASS detections \citep{ber96} but only covering 1\% of the entire sky (to be compared with the full sky investigated in the course of the RASS).

The derived properties of the hot stars confirm the results of the first preliminary data of the RASS and the recent in-depth investigations of a few clusters. The O stars have relatively soft spectra and exhibit a rather tight $L_{\rm X}-L_{\rm BOL}$ relation, although the dispersion is closer to that observed in the RASS survey than in a dedicated analysis of specific clusters. The B stars appear to have far harder spectra, with brighter emission for the earliest subtypes and a $L_{\rm X}-L_{\rm BOL}$ relation with far larger scatter. However, it must be noted that on average binaries do not appear significantly brighter than single objects.

The detection lists and the results of more than 300 spectral fits are made available to the community.

\begin{acknowledgements}
YN thanks Gregor Rauw for his careful reading of the manuscript. She acknowledges support from the Fonds National de la Recherche Scientifique (Belgium), the PRODEX XMM and Integral contracts, and the `Action de Recherche Concert\'ee' (CFWB-Acad\'emie Wallonie Europe).  ADS, CDS, and the XMM catalogs were used for preparing this document.
\end{acknowledgements}

\end{document}